\documentclass[english,english,letterpaper,pra,aps,superscriptaddress,floatfix,twocolumn,showpacs]{revtex4}
\usepackage{amsfonts}
\usepackage{pstricks}
\usepackage{pst-poly}
\usepackage{graphicx}
\usepackage{txfonts}
\usepackage{times}
\usepackage{multirow}
\usepackage{amssymb}
\usepackage{subeqnarray}

\newcommand{\beq}{\begin{equation}}
\newcommand{\eeq}{\end{equation}}
\newcommand{\beqa}{\begin{eqnarray}}
\newcommand{\eeqa}{\end{eqnarray}}
\newcommand{\beqar}{\begin{eqnarray*}}
\newcommand{\eeqar}{\end{eqnarray*}}

\def\bi{\begin{itemize}}
\def\ei{\end{itemize}}
\def\be{\begin{equation}}
\def\ee{\end{equation}}
\def\bea{\begin{eqnarray}}
\def\eea{\end{eqnarray}}
\def\ben{\begin{eqnarray*}}
\def\een{\end{eqnarray*}}

\def\>{\rangle}
\def\<{\langle}

\newcommand{\1} I 

\begin{document}

\title{Work measurement in an optomechanical quantum heat engine}
\author{Ying Dong}
\affiliation{B2 Institute, Department of Physics and College of Optical Sciences,
University of Arizona, Tucson, Arizona 85721, USA}
\affiliation{Department of Physics, Hangzhou Normal University, Hangzhou, Zhejiang 310036, China}

\author{Keye Zhang}
\affiliation{Quantum Institute for Light and Atoms, State Key Laboratory of Precision Spectroscopy, Department of Physics, East
China Normal University, Shanghai, 200241, China  }
\affiliation{B2 Institute, Department of Physics and College of Optical Sciences,
University of Arizona, Tucson, Arizona 85721, USA}
\author{Francesco Bariani}
\affiliation{B2 Institute, Department of Physics and College of Optical Sciences,
University of Arizona, Tucson, Arizona 85721, USA}

\author{Pierre Meystre}
\affiliation{B2 Institute, Department of Physics and College of Optical Sciences,
University of Arizona, Tucson, Arizona 85721, USA}

\begin{abstract}
We analyze theoretically the measurement of the mean output work and its fluctuations in a recently proposed optomechanical quantum heat engine [K. Zhang {\it et al.} Phys. Rev. Lett. {\bf112}, 150602 (2014)]. After showing that this work can be evaluated by a continuous measurements of the intracavity photon number we discuss both dispersive and absorptive measurement schemes and  analyze their back-action effects on the efficiency of the engine.  Both measurements are found to reduce the efficiency of the engine, but their back-action is both qualitatively and quantitatively different. For  dispersive measurements the efficiency decreases as a result of the mixing of photonic and phononic excitations, while for absorptive measurements, its reduction results from photon losses due to the interaction with the quantum probe.
\end{abstract}
\date{\today}

\maketitle

\section{Introduction}

The thermodynamic description of quantum heat engines (QHE) has been discussed at least since the early days of laser physics~\cite{Scovil1959} and has recently attracted much interest~\cite{Kosloff1992, Feldmann1996, Quan2007,kurizki2015}, in part because the increased control achievable over microscopic  and mesoscopic systems opens promising new avenues of theoretical and experimental investigation~\cite{Mazza2014,Liao2010,Venturelli2013,Dechant2014,Abah2012,Brunelli2014}.

QHE can exhibit intriguing properties, including their potential to outperform their classical analogues. For example, it has been shown that a quantum photo-Carnot engine can extract work from a single reservoir if the latter has built-in quantum coherence~\cite{Scully2003}, and its power can be increased by noise-induced coherence \cite{Scully2011}. In a different situation, a trapped ion based quantum engine operating on an Otto cycle was shown theoretically to break the Carnot efficiency limit in the presence of a squeezed reservoir~\cite{Abah2012}.

The definition of thermodynamical quantities in the quantum context presents however conceptual challenges~\cite{Allahverdyan2005, Esposito2006, Boukobza2006, Talkner2007}, and much attention has been devoted to the proper definition and the quantum statistical properties of quantities such as heat, work and entropy \cite{Fusco2014,Dorner2012,Mascarenhas2014,Tony2014,Campisi2013,Joshi2013,Smacchia2013,Saira2012,Batalhao2014,An2015,suomela2015}.  In closed quantum systems work may be defined in terms of a two-time measurement scheme~\cite{Esposito2009,Campisi2011,Hanggi07-11,Mukamel2003} or, in a recently proposed  alternative approach, of a single projective measurement~\cite{Roncaglia2014}. However the situation is less clear for open quantum systems, where there are still open questions regarding the definition and and experimental measurements of work and heat \cite{Esposito2009,Campisi2011,Crooks2008,Talkner09-10,Ritort2009} due to the lack of energy conservation in the reservoir(s). In this context quantum stochastic thermodynamics~\cite{Horowitz2012}, like its classical counterpart~\cite{Seifert05-08,Esposito07-10}, offers an interesting framework to discuss  thermodynamic properties  and simulate numerically the system behavior.

Optomechanical systems are prime candidates to investigate the properties of QHE. Thanks in particular to remarkable advances in nanofabrication they have witnessed rapid developments in the last decade and can now operate routinely deep in the quantum regime, with broad potential for applications in quantum technology~\cite{OMrev}. Recently,  three of us proposed and analyzed theoretically an optomechanical QHE based on an Otto cycle \cite{Zhang14}.  In this system the intracavity field of an optomechanical resonator interacts coherently with a single mode of vibration of a mechanical resonator. The basic idea  behind the thermodynamic cycle is that depending on the detuning between the driving optical field and the resonator the nature of the normal modes of the system (polaritons) can be changed from photon-like to phonon-like, with associated coupling to thermal reservoirs of different temperatures (cold for the photons and warmer for the phonons.) The exquisite experimental control that can be achieved in optomechanics suggests that such a QHE may be a good candidate to implement a measurement of the work output.

We consider that specific system to discuss several aspects of the work that can be extracted from QHE. Particular emphasis is placed on the quantum measurement of the work and its fluctuations, and also on the back-action of its measurement on the efficiency of the system. We first show that for that specific QHE the work can be evaluated from measurements of the  intracavity photon number. We consider and contrast an absorptive and a dispersive measurement scheme, both involving passing a stream of two-state atoms  through the resonator. The former situation results in projective measurements of the photon number, and the associated coupling between the normal modes of the optomechanical system, while the latter corresponds to the addition of an additional energy dissipation channel for the photons. We numerically determine the mean work and its variance over the entire thermodynamical cycle for both measurement schemes and use these results to evaluate the measurement back-action in the thermodynamic cycle. Our analysis is carried out within the framework of quantum stochastic thermodynamics, with the measured work evaluated via continuous detection of the mean photon number in the cavity which is responsible for the radiation pressure acting on the mechanical resonator.

The  paper is organized as follows. Section II briefly reviews the optomechanical QHE of Ref.~\cite{Zhang14} and the main features of the Otto cycle. In particular we draw attention to the fact that the two normal modes of the system undergo two distinct thermodynamic cycles. This will be important to keep in mind in the context of the measurement back-action of dispersive continuous measurements. Section III defines the work output of the engine, using the conceptually simpler case of classical measurements to show the relationship between the extracted work and the mean intracavity photon number. This result is used to justify the use of continuous measurements of the intracavity photon number operator to determine the expectation value of the work and its fluctuations in the quantum regime. Section IV introduces two specific types of continuous measurements that involve either the dispersive or the absorptive interaction between the cavity mode and a stream of two-state systems. Information on the intracavity field is then inferred from measurements of the state of the atoms as they exit the optomechanical resonator. Section V summarizes the results of numerical simulations obtained by a standard quantum trajectory approach to the solution of the stochastic master equations describing  the continuous measurements. Finally Section VI is a summary and outlook.

\section{Optomechanical Otto cycle}
This section briefly reviews the main features of the optomechanical quantum heat engine of Ref.~\cite{Zhang14}. We consider a standard optomechanical setup with a cavity mode of frequency $\omega_c$ and damping rate $\kappa$ coupled via radiation pressure to a single oscillation mode of a mechanical resonator of frequency $\omega_m$ and damping rate $\gamma$. The cavity is driven by an optical pump field of strength $\alpha_{\rm in}$ and frequency $\omega_p$.  We assume that the system reaches a steady state with mean intracavity field amplitude $\alpha \approx \alpha_{\rm in}/\Delta$, where
\begin{equation}
|\Delta| = |\omega_p-\omega_c - 2\beta g| \gg \kappa
\label{Eq:delta}
\end{equation}
is the detuning between the pump and cavity fields corrected for the equilibrium position of the mechanical oscillator, $\beta = -g\alpha^2/\omega_m$, and $g$ is the single-photon optomechanical coupling constant.

Denoting the small fluctuations of the photon and phonon modes around the steady state by the bosonic annihilation operators $\hat{a}$ and $\hat{b}$ respectively, the Hamiltonian for these fluctuations is
\begin{equation}\label{eq:H}
\hat H_{ab}=-\hbar\Delta\hat{a}^\dag\hat{a}+\hbar\omega_m\hat{b}^\dag\hat{b}+\hbar G(\hat{a}^\dag+\hat{a})(\hat{b}^\dag+\hat{b}), \label{eq:Hab}
\end{equation}
where we have introduced the linearized optomechanical coupling strength $G=\alpha g$.

We focus on the red detuned regime $\Delta < 0$, which in general leads to stable dynamics for small damping~\cite{OMrev} and perform a Bogoliubov transformation to diagonalize the Hamiltonian~(\ref{eq:Hab}) in terms of normal modes (polaritons) described by the bosonic annihilation operators $\hat{A}$ and $\hat{B}$. Ignoring a constant term that does not affect the dynamics, this gives
\begin{equation}\label{eq:HAB}
\hat H_{AB} =\hbar\omega_A\hat{A}^\dag\hat{A}+\hbar\omega_B\hat{B}^\dag\hat{B}, \label{eq:HAB}
\end{equation}
with normal mode eigenfrequencies
\begin{equation}
\omega_{A,B}=\sqrt{\frac{\Delta^2+\omega_m^2\pm\sqrt{(\Delta^2-\omega_m^2)-16G^2\Delta\omega_m}}{2}}.
\end{equation}
They are plotted in Fig.~1 as a function of the detuning $\Delta$.
It is straightforward to see that for $\Delta \ll -\omega_m$, the polariton ``A'' is photon-like and the polariton ``B'' is phonon-like, while in the opposite limit $-\omega_m \ll \Delta < 0$, it is the polariton ``A" that is phonon-like and the polariton ``B" is photon-like.

Consider then a situation where the phonon reservoir is at some finite temperature $T_{\rm phonon}$, while the optical field is coupled to a reservoir at $T=0$ -- an excellent approximation at visible frequencies -- and concentrate first on the polariton ``B" only. It is possible to realize an Otto cycle for that normal mode in the following way~\cite{qotto}: Start from the system in thermal equilibrium at a large negative detuning $\Delta$, in which case ``B" is essentially at the temperature of the phonon bath, with corresponding thermal excitation number $\langle\hat{B}^{\dagger}\hat{B}\rangle \equiv \bar N_B$, and adiabatically change $\Delta$ across the resonance $\Delta = -\omega_m$ and toward small negative value close to $0$ (to avoid the onset of instabilities).  In that first adiabatic stroke the polariton ``B" changes its character from phonon-like to photon-like, and the energy of the thermal phonons is converted into intracavity photons that perform work on the oscillating mirror via radiation pressure -- more on that in the following section. Once the detuning has reached its final value the system is then allowed to thermalize with the cavity field reservoir at $T=0$ (first thermalization stroke), releasing heat in the process. The third stroke is again adiabatic (second adiabatic stroke). It consists in changing $\Delta$ back to its large negative value. Finally the  cycle is closed by allowing the polariton ``B", which has now regained its phonon-like character, to thermalize with its reservoir at $T_{\rm phonon}$ by absorbing heat (second thermalization stroke.) These four strokes are sketched schematically in Fig. 1(b).

The polariton ``A" simultaneously also goes through a thermodynamic cycle, with however significant differences. First, it is initially coupled to a reservoir at $T\approx 0$, so that the initial thermal polariton occupation is $\bar N_A\approx 0$. Second, the first thermalization stroke for the ``B" cycle, which takes a time of the order of a few $\kappa^{-1}$, is not long enough to also thermalize  the ``A" polariton provided that the optical damping rate is much faster than the mechanical damping rate, $\kappa \gg \gamma$, which is normally the case in optomechanical systems. Under these conditions the population of mode ``A" remains essentially unchanged and equal to zero, and the ``A" cycle does not produce any work (positive or negative.) That is, provided that the changes in detuning $\Delta$ can be realized in a perfectly (quantum) adiabatic fashion, the two cycles remain completely decoupled. This, however, no longer holds if quantum adiabaticity cannot be maintained. This will have important consequences in the context of the dispersive quantum measurements of sections IV and V.

\begin{figure}[htbp]
\includegraphics[width=3.4in]{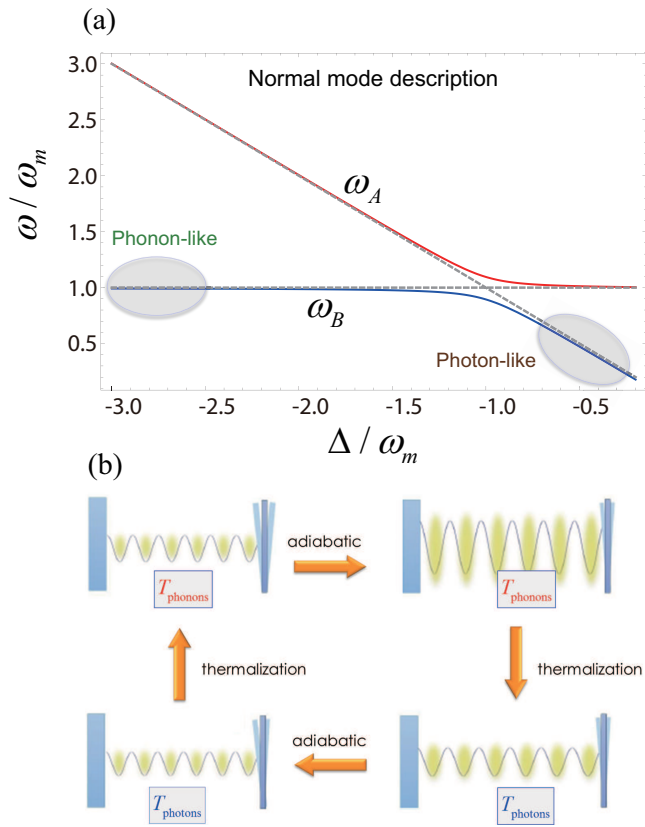}
\caption{(Color online) (a) Frequencies of the two normal modes (polaritons) of the optomechanical system for $G/\omega_m = 0.1$ in the red-detuned case $\Delta < 0$. The dashed lines correspond to the of the bare photon and phonon modes. The plot also indicates that for large negative detunings the ``B" polariton is phonon-like, and photon like for small negative detunings. (b) Sketch of the four strokes of the Otto cycle for the ``B" polariton. See text for details.}
\label{fig-fig1}
\end{figure}

\section{Output work}

We now turn to a discussion of the work performed by the optomechanical heat engine. It is useful to carry out this discussion both in the normal mode picture and the bare modes picture, if only because the measurements we have in mind will be on the optical field.

For notational convenience we decompose the energy operator formally as
\begin{equation}
\hat U = \hat W + \hat Q.
\end{equation}
Although there is still considerable debate in the literature about the nature of a work operator, as mentioned in the Introduction, this issue needs not concern us here as we are only concerned about operational ways to evaluate its first moment and fluctuations in a specific experimental setting.

The first law of thermodynamics, cast in infinitesimal form can then be formally written as
\begin{equation}
d \langle \hat U \rangle = d \langle \hat W\rangle + d \langle \hat Q \rangle,\label{eq:1st}
\end{equation}
where $\langle \hat U\rangle $, $\langle \hat Q \rangle $, and $\langle \hat W \rangle $, are the mean energy, heat, and work, respectively. For a general open quantum system with density operator $\rho$ and Hamiltonian $H$ we have
\begin{equation}
\langle \hat U \rangle = {\rm Tr}[\hat \rho \hat H],\label{eq:dU}
\end{equation}
so that in the Schr\"odinger representation the average (classical) values of the infinitesimal work and heat increments are~\cite{firstlaw}

\begin{eqnarray}
 d \langle \hat W \rangle  & = & {\rm Tr}\left [\hat \rho (d \hat H)\right ],\label{eq:dW}\\
d \langle \hat Q \rangle & = & {\rm Tr}\left [(d\hat \rho) \hat H\right ].\label{eq:dQ}
\end{eqnarray}

Our main focus is the measurement of the work produced by the system. For an isolated quantum system, that work may be determined unambiguously via a two-time measurement process~\cite{Esposito2009,Campisi2011}. However this approach is problematic for an open quantum system since it would require additional measurements on the reservoirs~\cite{Horowitz2012}. We therefore adopt an operational approach based on stochastic quantum thermodynamics~\cite{Horowitz2012}.

Specifically, our starting point is the description of the evolution of open quantum systems in terms of a large ensemble of $N$ quantum trajectories $\{|\psi_j(t)\rangle\}$ that are solutions of a stochastic Schr\"odinger equation of the general form
\begin{equation}
d|\psi(t)\rangle=(\mathcal{D}dt+\mathcal{R}dw)|\psi(t)\rangle,
\end{equation}
where the superoperator $\mathcal{D}dt$ accounts for both the Hamiltonian evolution of the system and non-unitary contributions that account for dissipation and decoherence mechanisms associated with measurement processes, and the stochastic term $\mathcal{R} dw$, where $dw$ describes one or more Wiener process of zero mean with $dw^2=dt$, accounts for the stochastic quantum jumps resulting from both reservoir noise and/or quantum measurement ~\cite{Carmichael1993,Wiseman1994,Plenio1998}. The initial conditions $|\psi_j(0)\rangle$ are randomly selected consistently with the initial density operator of the system.

The properties of any quantum observable can be calculated based upon the statistics deriving from these trajectories. In particular for each trajectory, we can compute the work by integrating the Eq. (\ref{eq:dW}) as
\begin{equation}
W_j = \int_{t_i}^{t_f}\langle\psi_j(t)|\frac{\partial \hat H}{\partial t}|\psi_j(t)\rangle dt
\end{equation}
where $t_i$ and $t_f$ are the initial and final times. The resulting mean value and the variance of the work are then obtained as
\begin{equation}
\langle \hat W \rangle= \sum_{j=1}^{N} \frac{W_j}{N}, \label{Wbar}
\end{equation}
and
\begin{equation}
\Delta W^2= \sum_{j=1}^{N} \frac{(W_j-\langle \hat W \rangle)^2}{N} = \langle \hat W^2 \rangle - \langle \hat W \rangle^2. \label{Wvar}
\end{equation}
In the limit $N\rightarrow \infty$, the statistics resulting from the quantum trajectories approach the correct result. The numerical simulations presented in section V are based on that approach.

\subsection{Output work in the optomechanical QHE}

We have seen that in the normal mode picture the ``B" polariton heat engine is driven through an Otto cycle. Provided that dissipation is weak enough to be negligible during the adiabatic strokes only work is performed during those  strokes, while heat is only exchanged during the thermalization steps~\cite{Zhang14}. Since the ``B" polariton population is $\bar N_B=0$ after thermalization with the optical heat bath at $T=0$ we can then restrict the determination of the output work to the first adiabatic stroke, where
\begin{equation}
d\langle \hat Q \rangle ={\rm Tr}[(d\hat \rho_{AB}) \hat H_{AB}]=0,
\end{equation}
and
\begin{equation}
d \langle \hat W \rangle ={\rm Tr}[\hat \rho_{AB}( d \hat H_{AB})]= \bar{N}_B \hbar d\omega_B. \label{eq:W1}
\end{equation}
Here we have used the polariton Hamiltonian (\ref{eq:HAB}) and assumed that $\bar N_A\approx 0$, as previously discussed.  Since in the adiabatic stoke $\bar N_B$ is conserved, the average work is simply given by the change in energy of mode ``B",
\begin{equation}
\langle \hat W \rangle  =\int_{\Delta_{i}}^{\Delta_{f}} dW=  \hbar[\omega_{B}(\Delta_{f})-\omega_{B}(\Delta_{i})]\bar{N}_{B}, \label{eq:W1}
\end{equation}
where $\Delta_i$ and $\Delta_f$ are the initial and final detunings, see Eq.~(\ref{Eq:delta}). Note also that in the case of perfect adiabaticity, the ``B" polariton number distribution gives  directly the full statistical distribution of the work performed by the system as well.

In the bare mode representation, the first stroke is not adiabatic, and the photon and phonon distributions are time dependent. Equation~(\ref{eq:dW}) then takes the form
\begin{equation}
 d\langle\hat W \rangle ={\rm Tr}[\hat \rho_{ab} (d \hat H_{ab})]=-\bar n_a \hbar d\Delta,
\end{equation}
where we used the Hamiltonian (\ref{eq:Hab}) and $\bar n_a=\langle \hat a^\dagger \hat a \rangle$ is the average number of excitations in the photon mode $\hat{a}$. In this picture the work is then given by
\begin{equation}
\langle \hat W \rangle=-\hbar \int_{\Delta_{i}}^{\Delta_{f}} \bar n_a (\Delta) d\Delta. \label{eq:Wa}
\end{equation}
If the stroke is perfectly adiabatic the values of the average work obtained from expressions (\ref{eq:W1}) and (\ref{eq:Wa}) are equal. However if either the optical or the mechanical damping is significant on the time scale of the adiabatic stroke, or if the variation of the optical detuning induces non-adiabatic transitions and in particular a non-vanishing population of polariton $\hat{A}$, then Eq.~~(\ref{eq:W1}) is no longer exact. The expression of the average work~(\ref{eq:Wa}) in terms of the mean photon number remains however valid.

\section{QHE work measurement}

We now discuss several possible measurement schemes that can be considered to quantify the work performed by the heat engine and its fluctuations. To set the stage we first consider a simple classical approach before considering two types of quantum measurements.

\subsection{Classical measurement scheme}

We can think of implementing the variation in detuning $\Delta$ required for the adiabatic stroke through a change in cavity length
\begin{equation}
\Delta(y)=\Delta_0-g_M y,	
\end{equation}
where $y$ is a {\em classically} controlled length change, assumed small with respect to the total cavity length, $g_M\equiv g/y_{M}$ is the optomechanical coupling normalized to the mirror zero-point motion $y_M$ , and $\Delta_0$ is the initial detuning. With Eq.~(\ref{eq:Wa}) we can then express the work in terms of the spatial integral of the position-dependent radiation pressure force,
\begin{equation}
\langle \hat W \rangle = \int_{y_i}^{y_f} F_{\rm rp}(y) dy,\label{eq:W2}
\end{equation}
where
\begin{equation}
F_{\rm rp}(y) = \hbar g_M \bar n_a(y).
\end{equation}

\begin{figure}[htbp]
\includegraphics[width=3.4in]{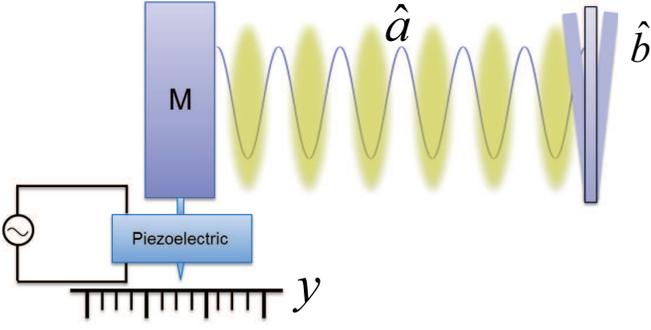}
\caption{(Color online) Schematic setup for the classical measurement of the output work, whereby the work performed by the radiation pressure acting on the mirror of large mass $M$ can be stored in the control system. See text for details.}
\label{fig-fig2}
\end{figure}

A possible classical scheme to measure the work output of the engine is illlustrated in Fig.~\ref{fig-fig2}. The optomechanical resonator comprises the oscillating end mirror driven by radiation pressure and a mirror of large mass $M$ whose {\em classical} position can be controlled externally by the potential $V(y)$ provided by a piezoelectric element, therby varying the detuning $\Delta(y)$ in the presence of the radiation force $F_{\rm rp}$.  To use a thermodynamical metaphor, we may think of  the input mirror as a classical piston that is pushed by the expanding photon gas.

The total system Hamiltonian is then
\begin{equation}
\hat H_T= \hat H_{ab} + H_M,\label{eq:Hom}
\end{equation}
where $H_{ab}$ is given by Eq.~(\ref{eq:Hab}) and
\begin{equation}
H_M=\frac{p^{2}}{2M}+V(y), \label{eq:HM}
\end{equation}
is the classical Hamiltonian for the massive control mirror. The classical equations of motion for that mirror are then
\begin{eqnarray}
\frac{dy}{dt} & = & \frac{\partial H_T}{\partial p}=\frac{p}{M},\\
\frac{dp}{dt} & = & -\frac{\partial H_T}{\partial y}=-\frac{\partial V(y)}{\partial y} - F_{\rm rp},
\end{eqnarray}
where $H_T$ is the classical limit of the total Hamiltonian $\hat H_T$. If $M$ is large enough that it can be considered as infinite compared to all other optomechanical elements we have  $dy/dt \approx dp/dt\approx 0$. That is, the force exerted by the control system balances the expectation value of the radiation pressure force,
\begin{equation}
-\frac{\partial V(y)}{\partial y}=  F_{\rm rp},
\end{equation}
This shows that provided the kinetic energy of the large mirror remains essentially zero all work performed by the photons is converted to the control potential energy and can be measured in that way.

\subsection{Continuous quantum measurements}
It is not possible to directly monitor the occupation of the polariton mode ``B" since it consists of  quasiparticles that are coherent superpositions of photon and phonon states. Instead, and in analogy with the classical measurement scheme we apply a weak continuous measurement scheme~\cite{Jacobs2006} to monitor the intracavity photon number and calculate the total work by performing the integral in Eq.~(\ref{eq:Wa}). Similarly to the classical scheme the total system Hamiltonian is then of the form
\begin{equation}
\hat H_T=\hat H_{ab}+\hat V_m,
\end{equation}
with $\hat V_m$ describing the weak interaction between the photon and the measuring quantum system. Since in general $\hat V_m$ does not commute with the optomechanical Hamiltonian $\hat H_{ab}$ the measurements lead in general to back-action on the QHE that affects output work and its efficiency. In the following we will use the method of quantum trajectories to simulate the measurement processes and investigate their influence on the mean work $\langle \hat W\rangle $ and its fluctuation $\Delta W^2$.

\begin{figure}
\includegraphics[width=3.4in]{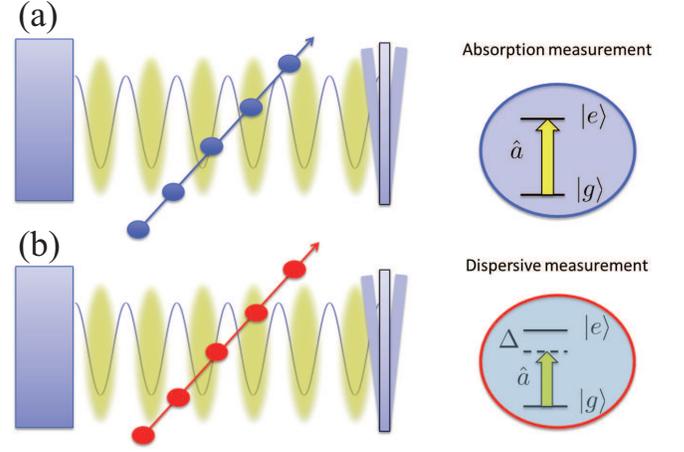}
\caption{(Color online) Schematic setup for a continuous quantum measurement of the output work with a beam of two-level atoms. a) Absorptive measurement: the cavity field is resonant with the atomic transition and the coupling induces real oscillations in the atomic population, which result in the loss of intracavity photons. b) Dispersive measurement: the cavity mode frequency is far off-resonant from the two-level atom transition frequency, resulting in a dispersive interaction that only modifies the phase of the atomic ground state wave function. See text for details.}
\label{fig-fig3}
\end{figure}

Operationally we consider continuous measurements of the intracavity field realized by passing through the resonator a dilute beam of two-level atoms that interact weakly with it, with at most one atom at a time inside the resonator. The state of the field is then inferred from a measurement on the atoms after they exit the cavity. We study both the cases of absorptive and dispersive atom-field inteactions, see Fig.~\ref{fig-fig3}.

\subsubsection{ Absorptive measurements}

Consider first the resonant situation where the atomic transition frequency $\omega_{eg} = \omega_c$, and the atoms are prepared in their ground state $|g\rangle$ before being injected inside the optical cavity, see Fig.~3(a). The atom-field coupling is given in the rotating wave approximation by
\begin{equation}
\hat V_m =  \hbar g_a (\hat{a}^\dag\hat{\sigma}_{ge} + \hat{a}\hat{\sigma}_{eg})
\label{Eq:Va}
\end{equation}
where $g_a$ is the single-photon Rabi frequency of the transition and $\hat{\sigma}_{ij}= | i \rangle\langle j | $, $j=\{e,g\}$. The effect of the continuous measurements on the reduced density operator for the optical field is then governed by the stochastic master equation~\cite{Imoto1990,Jacobs2006}
\begin{eqnarray}
d\hat{\rho}(t)&=& \frac12 \lambda_a \left (2\hat{a}\hat{\rho}\hat{a}^\dag-\hat{a}^\dag\hat{a}\hat{\rho}-\hat{\rho}\hat{a}^\dag\hat{a}\right )dt\nonumber \\
&+&\sqrt{\lambda_a}\left (\hat{a}\hat{\rho}+\hat{\rho}\hat{a}-2\langle \hat a + \hat a^\dagger\rangle\hat{\rho} \right ) dw
\label{Eq:me1}
\end{eqnarray}
where $dw$ is a Wiener process and
\begin{equation}
\lambda_a = g_a^2 \tau
\end{equation}
is a measure of the strength of the measurement, $\tau$ being the transit time of an individual atom through the resonator. Note that in obtaining Eq.~(\ref{Eq:me1}) the time increment $dt$ is assumed to be long compared to the atomic transit time $\tau$, so that this equation describes the statistical effect on the field of a large number of atomic measurements.

The first term on the right-hand side of Eq.~(\ref{Eq:me1}) accounts for the additional dissipation channel of the intracavity field resulting from absorption by the successive atoms, and the second term describes the stochastic changes on the intracavity field about its expected value $\langle \hat a + \hat a^\dagger\rangle$ as a result of the measurement outcomes.

\subsubsection{Dispersive measurements}

We now turn to the situation where the interaction between the two-level atoms and the intracavity field mode is off-resonant. Upon adiabatic elimination of the upper electronic stateit is described by the effective Hamiltonian
\begin{equation}
\hat V_m=\hbar g_d\hat{a}^\dag \hat{a}(\hat{\sigma}_{ee}-\hat{\sigma}_{gg})=\hbar g_d\hat{a}^\dag \hat{a}(\hat{\sigma}_{+-} + \hat{\sigma}_{-+}),
\label{Eq:Vd}
\end{equation}
which conserves the mean photon number $\bar n_a=\langle \hat a^\dagger \hat a\rangle$. Here $g_d = g_a^2/2\delta$ is the off-resonant coupling between the intensity of the field and the energy of the atomic levels.

The atoms are now prepared in the superposition $|+\rangle = (|e\rangle + |g\rangle)/\sqrt{2}$ and information on the intracavity field is inferred from a change in phase of the atomic state. In that situation the effect of the measurements on the optical field  is described by the stochastic master equation~\cite{Ueda1992}
\begin{eqnarray}
d \hat{\rho}(t)& = & \frac12 \lambda_d (2\hat{n}_a\hat{\rho}\hat{n}_a-\hat{n}_a^2\hat{\rho}-\hat{\rho}\hat{n}_a^2)dt 
\nonumber \\
&&+\sqrt{\lambda_d}(\hat{n}_a\hat{\rho}+\hat{\rho}\hat{n}_a-2 \langle \hat{n}_a\rangle\hat{\rho}) dw,
\label{Eq:me2}
\end{eqnarray}
where $\lambda_d=g_d^2\tau$.

As was the case for resonant coupling this equation also comprises two contributions, the second one accounting for the stochastic changes of the mean intracavity intensity $\langle \hat n\rangle$ about its expected value resulting from successive measurements. But because of the quantum non-demolition nature of the non-resonant atom-field interaction, the dissipative channel of Eq.~(\ref{Eq:me1}) is now replaced by a number conserving term that results in an additional damping of the phase of the optical field.

Importantly, in the specific case of our optomechanical QHE the effective interaction~(\ref{Eq:Vd}) couples the two polariton branches ``A" and ``B", transferring excitations between these two modes. As we see in section V this can be thought of as a nonadiabatic coupling that has in general a significant effect on the work that can be extracted in the Otto cycle of the ``B" polariton~\cite{Zhang14}. This is in contrast to absorptive measurements, as the interaction~(\ref{Eq:Va}), does not significantly couple the two normal mode branches. Still, both measurement schemes result in the appearance of additional photon loss channels that limit the amount of extractable work. In both cases these measurement back-action mechanisms may be viewed as heat exchange between the engine and the environment.

\section{Numerical results}
This section presents selected results from numerical simulations of the continuous measurement of the work output of the QHE and its fluctuations as defined in equations (\ref{Wbar}) and (\ref{Wvar}), both for absorptive and dispersive measurements. The numerical results were obtained by averaging for each choice of parameters 20,000 trajectories obtained from a combination of the Hamiltonian evolution of the system as the detuning $\Delta(t)$ is varied across the Otto cycle and the solution of the stochastic Schr\"odinger equations
\begin{eqnarray}
d|\psi\rangle&=&\left \{ \left [-\frac 12 \lambda_d \left (\hat{n}_a-\langle \hat n_a\rangle\right )^2\right ]dt \right.
\nonumber \\
&&+\left. \sqrt{\lambda_d}(\hat{n}_a-\langle \hat n_a\rangle)dw\right \}|\psi(t)\rangle,
\end{eqnarray}
and
\begin{eqnarray}
d|\psi\rangle&=&\left \{\left [-\frac12 \lambda_a (\hat{a}^\dag\hat{a}-\langle \hat a + \hat a^\dagger\rangle\hat{a}+\frac{\langle \hat a + \hat a^\dagger\rangle^2}{4})\right ]dt \right . \nonumber \\
&&+\left . \sqrt{\lambda_a}(\hat{a}-\frac{\langle \hat a + \hat a^\dagger\rangle}{2})dw  \right \}|\psi(t)\rangle,
\end{eqnarray}
corresponding to the stochastic master equations~(\ref{Eq:me1}) and (\ref{Eq:me2}) to account for the continuous measurements.

\begin{figure}
\includegraphics[width=3.4in]{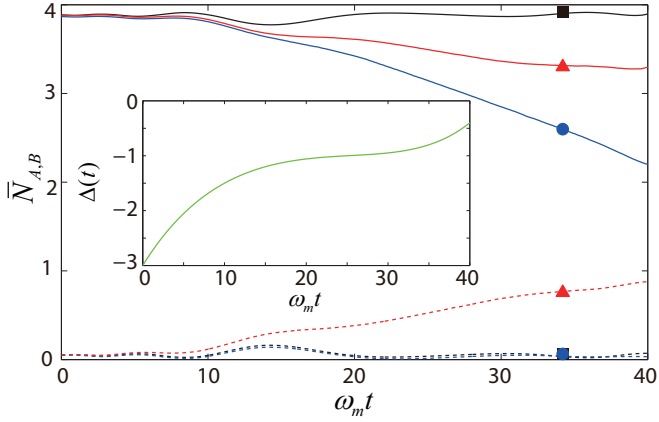}
\caption{(Color online) Time evolution of the mean excitations $\bar N_B$ and $\bar N_A$ of the polariton modes  ``B" (solid lines) and ``A" (dashed lines) during the first stroke of the heat engine, averaged over 20,000 trajectories of the stochastic Schr\"odinger equation. Black lines marked by squares: no measurement.  Red lines marked by triangles: Dispersive measurement with $\lambda_d=0.04\omega_m$. Blue lines with circles: absorptive measurements with $\lambda_a=0.04\omega_m$. Other parameters: $G=0.2\omega_m$, $\Delta_i=-3\omega_m$, $\Delta_f=-0.4\omega_m$ ,$\kappa=5\times10^{-3}\omega_m$ and $\gamma=10^{-4}\omega_m$. Inset:  time dependence of the pump-cavity detuning $\Delta(t)$ in units of $\omega_m$. Time in units of $1/\omega_m$.}
\label{fig-fig4}
\end{figure}

To guarantee that the adiabatic strokes are indeed adiabatic in the absence of measurements, it is important to change $\Delta(t)$ sufficiently slowly that non-adiabatic transitions between the two polariton branches remain negligible, but fast enough that the damping of both the optical field and the mechanical oscillator at rates $\kappa$ and $\gamma$ respectively,  remain negligible. This is particularly the case near the avoided crossing at $\Delta = -\omega_m$. The insert of Fig.~\ref{fig-fig4} shows as an example the time evolution of $\Delta(t)$ (in units of $\omega_m$) used in the simulations of the first adiabatic stroke to minimize this problem, with $\Delta(t)$ changing rapidly away from the avoided crossing and very slowly in its vicinity. As a result non-adiabatic transitions remain negligible, as illustrated in the black curves (labeled by a square) of Fig.~\ref{fig-fig4}. The population of the ``B" polariton mode remains essentially constant during the adiabatic stroke, and the ``A" polariton population remains essentially zero.

The additional curves in Fig. \ref{fig-fig4} show the time dependence of the average populations $\bar{N}_A(t)$ and $\bar N_B(t)$ during the first stroke of the heat engine, again neglecting mechanical and optical damping, for both dispersive (red lines with triangles) and absorptive (blue lines with circles) measurements. (Note that $\bar N_A$ remains extremely small during that stroke and its evolution is nearly indistinguishable from the situation without measurements.) This is the most important stroke as far as extracting work from the engine is concerned, since the population of normal mode ``B" remains extremely small during the second adiabatic stroke as a result of its thermalization at the optical reservoir temperature $T=0$. In these examples the temperature $T_{\rm phonon}$ of the phonon bath and the initial detuning $\Delta(0)$ are chosen such that $\bar{N}_B(0) \approx 4$ and $\bar{N}_A(0) \approx 0$.

Comparing the evolution of the mean populations of the polariton modes for absorptive and dispersive measurements illustrates clearly the difference in their back-action on the operation of the QHE. In the case of absorptive measurements the ``B" polariton population decreases significantly during what would otherwise be an adiabatic, population-conserving stroke. Since that stroke occurs fast compared to $\kappa^{-1}$ that damping results solely from an additional photon dissipation at rate $\lambda_a$ that results from the measurements, see Eq.~(\ref{Eq:me1}). Importantly, though, these measurements do not result in any significant transfer of population to the ``A" polariton.

The situation is qualitatively quite different for dispersive measurements. In that case there is no significant loss in total polariton population, but instead a significant transfer of population from mode ``B" to mode ``A". This is because the effect of dispersive measurements is an additional source of decoherence, but no loss of population, see Eq.~(\ref{Eq:me2}). Dispersive measurements change the frequency of the photons stochastically, as seen by the term proportional to $\hat a^\dagger \hat a$ in the master equation (\ref{Eq:me2}), and thereby they change the structure of the polaritons. In contrast, in the absorptive case the measurements remove excitations from the system, but without affecting the structure of the polaritons.

The population transfer between the normal modes ``B" and ``A" associated with dispersive measurements causes a reduction in the work performed by the system that can become quite dramatic due to the resulting unavoidable coupling between the two normal modes. Since for the polariton branch ``A" the Otto cycle is reversed and produces positive work~\cite{Zhang14}, that is, work is performed by the environment on the polariton~\cite{negwork}, one can even reach situations where the effective available work of the two systems, which are inextricably coupled,  becomes positive. For absorptive measurement, in contrast, the two normal modes remain essentially uncoupled, and although the output work can be significantly reduced it always remains negative.

\begin{figure}
\includegraphics[width=3.4in]{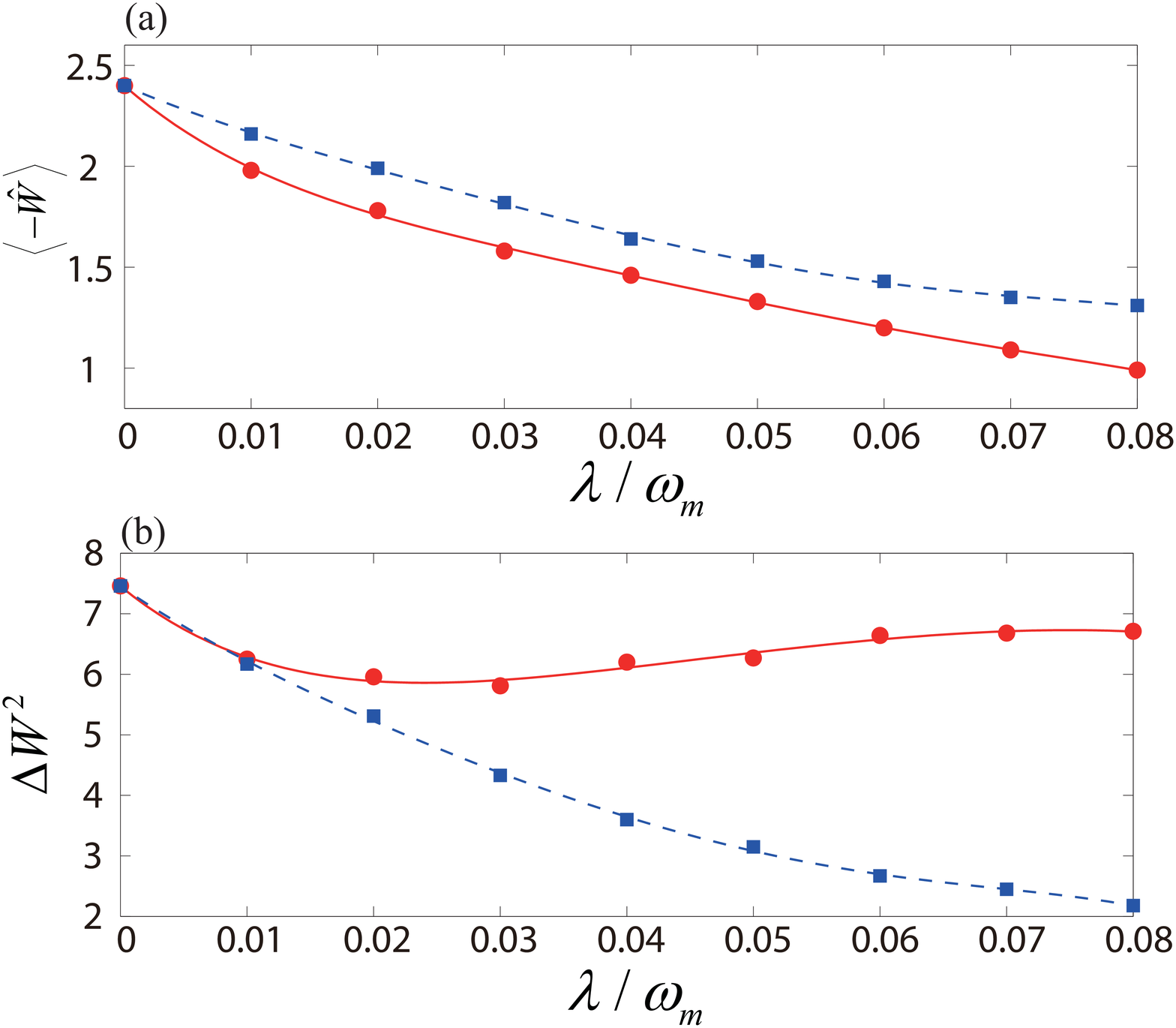}
\caption{(Color online) (a) Expectation value and (b) variance of the output work , in units of $\hbar \omega_m$, for the full cycle as a function of the measurement strenght (in units of $\omega_m$). The points are results of numerical simulations and the lines serve to guid the eye. In both figures, the sqaures (red solid line) and circles ( blue dashed line) stand for the dispersive and absorptive measurement scheme respectively. The statistic is done upon 20000 trajectories. The stroke times are $t_1=40\omega^{-1}_m$, $t=400\omega^{-1}_m$, $t_3=40\omega^{-1}_m$, $t_4=4\times10^4\omega^{-1}_m$. All other parameters as in Fig.~\ref{fig-fig4}.}
\label{fig-fig5}
\end{figure}

Figure~\ref{fig-fig5}(a) shows the mean value of the output work $-\langle \hat W \rangle$ for increasing measurement strength $\lambda_{a,d}$, illustrating its reduction due to measurement back-action. Surprisingly perhaps for equal measurement strengths dispersive measurements cause a stronger reduction in work than absorptive measurements. Since the thermalization processes are not affected by the measurement scheme the heat absorbed by the system from the mechanical reservoir, $\langle \hat Q_{\rm in}\rangle$, remains the same for all scenarios. For this reason, the efficiency of the quantum heat engine follows directly from the work as
\begin{equation}
\eta=\frac{\langle -\hat W \rangle  }{\langle \hat Q_{\rm in}\rangle}
\end{equation}
where the minus sign accounts for the fact that by convention the work done by the engine is negative ($\langle \hat W \rangle<0$).

We now turn to the fluctuations of the output work. Their variance, plotted in Fig.~\ref{fig-fig5}(b), shows a significant quantitative difference between the situations for absorptive and dispersive measurements. In the first case (dotted line) the fluctuations decrease monotonically as a function of the measurement strength, while in the case of dispersive measurements they remain roughly constant.

\begin{figure*}
\includegraphics[width=7in]{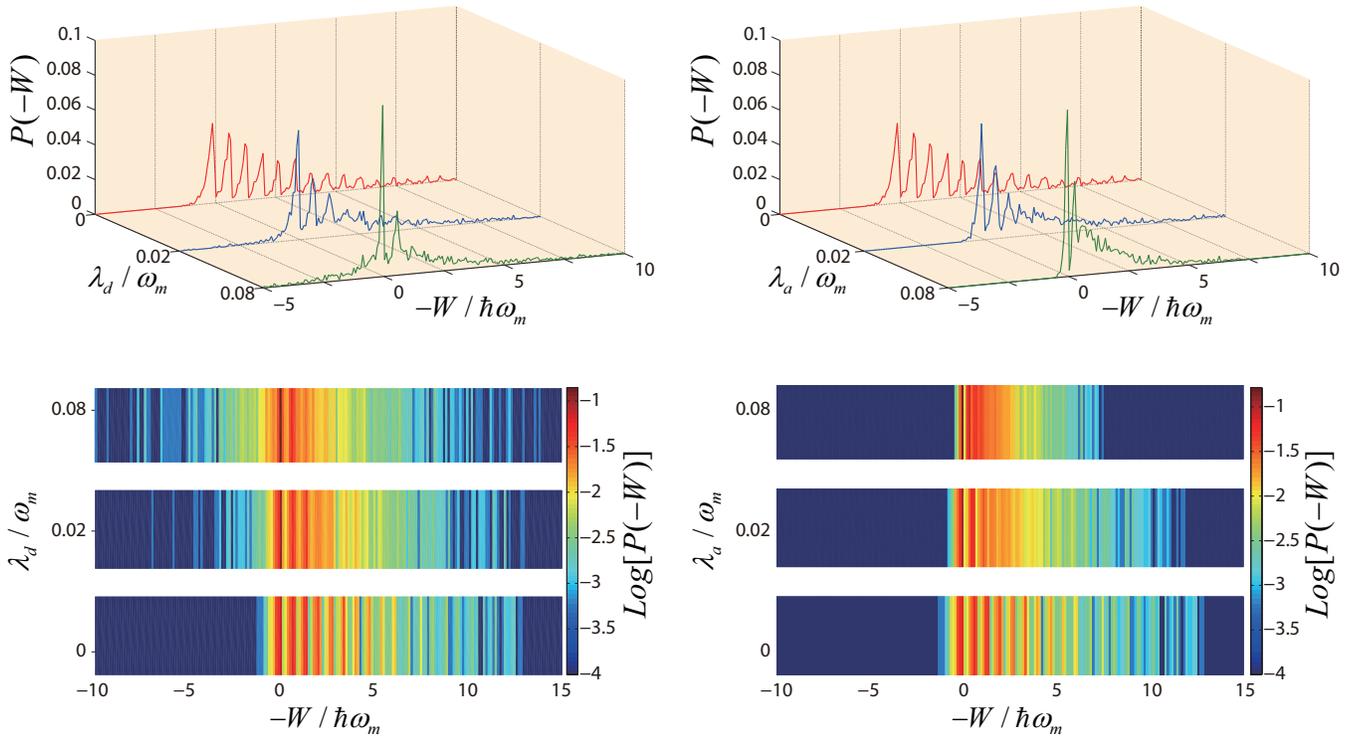}
\caption{Linear (upper panel) and logarithmic (lower panel) plots of the probability distribution of the output work, $\hat W$ (in units of $\omega_m$) for dispersive (left column) and absorptive(right column) measurement respectively in the absence of measuremnt and for two  measurement strength values. which label the axes All parameters are same as Fig.4. }
\label{fig-fig6}
\end{figure*}

One can gain a better understanding of this behavior from the probability distribution $P(-W)$ of the output work as a function of measurement strength. Figure~\ref{fig-fig6} shows this distributions in the absence of measurements and for two measurement strengths, for both dispersive and absorptive measurements. Without measurements the probability distribution consists of a series of discrete peaks that correspond to the distribution of Fock states in the initial thermal distribution of the mechanical oscillator. Assuming perfect adiabaticity each of these mechanical Fock states is converted into a photonic Fock state during the first adiabatic stroke, and produces a specific amount of work. (The width of the peaks is due to residual non-adiabatic effects.)

Continuous measurements result in a broadening of the peaks, an effect of the stochastic nature of the detection process (see upper panels of Fig.~\ref{fig-fig6}) and, in the case of absorptive measurements, a decrease in amplitude of all peaks except the one corresponding to the vacuum field, a consequence of the additional photon decay channel. This is the  reason for the reduction in variance as the measurement strength is increased.  In contrast, for dispersive measurements the distribution shifts toward positive values of the work. This is more apparent in the lower panels, which show the same distribution on a logarithmic scale. This is a direct consequence of the coupling with the ``A"-polariton engine which, as we have seen, tends to be characterized by positive work.  Because absorptive measurements don't couple the polariton modes in any significant way, this effect is almost completely absent in that case. Finally, since the mean total number of polaritons in modes ``A" and ``B" varies only slightly over the chosen measurement strengths, and the changes in the photon distribution are much less significant than for absorptive measurements, resulting is weak changes in the variance of the extracted work as a function of measurement strength.

\section{Summary and outlook}
Summarizing, we have developped a measurement model to characterize the mean work and its fluctuations in an optomechanical QHE and performed a numerical study of the effect of continuous quantum measurements on its performance. We considered measurement schemes involving the continuous monitoring of the intracavity photon field, with both dispersive and an absorptive interactions with a dilute beam of two-level atoms. By determining the average value and the variance of the work we are able to quantify the measurement back-action effects. In both cases, the measurements were found to induce a reduction in the average work performed by the engine and thus a reduction in its efficiency. However, the detailed reasons behind these reductions are qualitatively different. In the dispersive regime, the measurement induces transitions between the two polariton modes, and hence two thermodynamic cycles, one producing negative and the other positive work. The final result is a reduction of the efficiency with increased fluctuations in the work output. In the absorptive detection scenario, in contrast, photons are lost from the system via the interaction with the quantum probe that acts as an effective (zero temperature) reservoir. In this case, both the average value and the fluctuations of the work decay monotonically.

Perhaps the most intriguing result of this study is the realization that quantum measurements permit to control the operation of coupled QHE, and to switch the operation from a cycle transferring energy from a hot to a cold reservoir to the reverse situation. This suggests that it might be possible to consider quantum heat pumps whose operation is controlled by the back-action of quantum measurements. This and other aspects of QHE will be further explored in future work.

\acknowledgments
We acknowledge several enlightening discussions with A.A. Clerk. This work is supported by the DARPA QuASAR and ORCHID programs through grants from AFOSR and ARO, the U.S. Army Research Office, and NSF. YD is supported in part by the NSFC Grants No. 11304072 and the Hangzhou-city Quantum Information and Quantum Optics Innovation Research Team. KZ is supported by the NSFC Grants No. 11204084, No. 91436211, and No. 11234003, the National Basic Research Program of China Grant No. 2011CB921604, the SRFDP Grant No. 20120076120003, and the SCST Grant No. 12ZR1443400.


\begin{thebibliography}{99}
\bibitem{negwork}Remember that in the thermodynamic convention  work performed by the engine is negative.
\end{thebibliography}
\end{document}